# Data Management in Integrated Research Institutes: Undertaking a Review of Research Data Management at the Rosalind Franklin Institute

Felicity Currie[1,3], Mark Basham[2,4], Laura Shemilt[2] Nick Lynch[1],

Author Affiliations:

1. Curlew Research Limited, Buckinghamshire, UK
2. Artificial Intelligence and Informatics, The Rosalind Franklin Institute, Oxfordshire, UK

Corresponding Authors:

3. felicity.currie@curlewresearch.com
4. mark.basham@rfi.ac.uk

ORCID ID:

Felicity Currie https://orcid.org/0000-0002-3088-6223

Mark Basham https://orcid.org/0000-0002-8438-1415

Laura Shemilt https://orcid.org/0000-0001-5199-5624

Nick Lynch https://orcid.org/0000-0002-8997-5298

Author's contributions:

Felicity Currie collected user stories, ran the data review workshops, designed the survey, analysed results and wrote the manuscript

Mark Basham conceptualization of the review and contributed to the manuscript

Laura Shemilt conceptualization of the review and contributed to the manuscript

Nick Lynch collected user stories, ran the data review workshops, analysed results and contributed to the manuscript

## Abstract

Managing Research Data, and making it available for use/reuse by others in line with the UKRI Concordat on Open Research Data and FAIR principles, is a major issue for research-intensive organisations. In this case study we outline an institute-wide review of data management in practice, carried out at the Rosalind Franklin Institute (The Franklin) in partnership with external consultants, Curlew Research, in March 2022. We aim to describe the processes involved in undertaking a review of the services already in place that support good practice in managing research data, and their uptake, with an emphasis on the methodology used. We conducted interviews with scientific Theme Leads which set the scope for the Data Management Workshops subsequently held with researchers. Workshops were valuable in both providing actionable insights for data management and in priming the audience for future discussions. The final report produced for The Franklin, summarising results of the analysis, provides a snapshot of current practice for the Institute, highlighting what is working well and where improvements might be made, and provides a benchmark against which development can be measured in the coming years. The Review will continue to be conducted on an annual basis, reflecting changes in a fast-moving area and enabling an agile approach to research data management.

## Introduction

Data lies at the core of life sciences, biotech, across research institutes and pharmaceutical companies. But without thoughtful organisation and management of that data, the work of scientists and data analysts can be severely hampered, both within the organisation and for those collaborators, external partners and the wider scientific community seeking to use & reuse the data more widely. Equally, there is pressure from research funding bodies globally to support Open Data access (*e.g.,* UKRI (2023)) adopting FAIR principles (Wilkinson *et al.* 2016) and the wider key data management principles that underpin these activities.

In this article we describe the methodological approaches adopted by Curlew Research Ltd. in conducting an analysis of data management strategies. We present a use-case for a review conducted at the Rosalind Franklin Institute, who wished to review its data management approaches and associated platforms in the context of the UKRI Concordat on Open Research Data (UKRI 2016). The institute-wide analysis, carried out by Curlew Research, provided the Franklin with material to support future data governance planning and activities, and support both the needs of the key stakeholders within the Franklin, but also key collaborators, both internal and external.

The Rosalind Franklin Institute (The Franklin) is the UK's technology institute for life science. The Franklin was created by EPSRC in 2018, and was proposed by a group of ten leading UK Universities, who recognised the need for a dedicated resource to develop transformative technologies, beyond the scope of single or short-term grants. Franklin technologies aspire to have a 'Factor of 10' impact – with research projects looking for an order of magnitude change in analysis (*e.g.*, in speed, resolution) as an anticipated outcome from their creation. As such there is an accepted level of risk, with any research undertaken not necessarily subject to the same measures of success as other UKRI Institutes.

The Franklin currently has around 150 employees, student and secondees, rising to 250 over the next 2 years, and caters for visiting scientists and collaborators in academia and industry. The Franklin member community comprises the 10 founding UK Universities, Diamond Light Source and the UKRI Science and Technology Facilities Council (UKRI-STFC).

Research at the Franklin is structured along five Scientific Themes:

- Artificial Intelligence (AI) and Informatics
- Biological Mass Spectrometry
- Correlated Imaging
- Next Generation Chemistry
- Structural Biology

The Research Data Informatics Strategy is provided within the Artificial Intelligence and Informatics Theme in addition to AI research projects.

Research at the Franklin is wide-ranging, using analytical techniques that include Cryo-EM (sited both at the Franklin and also at the Diamond Light Source), Crystallography, Focused Ion Beam Scanning Electron Microscopy (FIB SEM), Confocal Microscopy, Mass spectrometry, NMR, and Robotics for synthetic chemistry. The Institute's Informatics Strategy covers 20 instruments (and >30 users). There is a distinction to be emphasised here between the larger instruments producing 'big data' (*e.g.,* for electron microscopy) and smaller instruments used in the 'wet lab' (*e.g.,* in the course of protein purification) with mainly proprietary or Excel-based data capture.

The Rosalind Franklin Institute is a new Institute, and, with no existing computing infrastructure, the decision was made to use existing and nearby compute and storage resources from Science and Technology Facilities Council (STFC) nearby facilities (*e.g.,*

STFC Ceph Echo Object store). This allows the Franklin to make the most of existing experience and services and essentially have a hybrid Cloud approach from day one of the Institute.

For instance, some examples of this partnering and use of existing infrastructure includes

- The Franklin makes use of open-source structure/software where possible and work with two partners:
  - STFC Open Stack Cloud for computation and services (STFC 2023)
  - STFC Ceph Echo Object Store for storage (Greeves 2020)
  - AWS where SFTC cannot meet the Franklin's needs
- For High Performance Computing (HPC) the Franklin partner with the Baskerville project at Birmingham (Baskerville 2021).

A number of key principles were adopted early on in developing the early data management approaches including:

- Aim to capture data early on in its lifecycle as possible
- Data can be ultimately downloaded to Franklin virtual machines (VMs) and external compute if authorised for data analysis / processing
- Focus on the data files from the instruments rather than attempting to interact with the instrument's software
  - All instruments run bespoke software & the Franklin AI team cannot control instrumental computers nor get easy access
  - Data files provide a decoupled approach to data

The initial focus for data capture was for the Correlated Imaging and Structural Biology themes given the large volumes of data and risk of data inaccessibility and possible loss if not tackled from start of operation.

The data management prioritisation approach led to the development of RFI File Monitor by the Data Team (Schoonjans 2021) and establishment of the Franklin instance of the SciCat data catalog (building on the existing SciCat Open Source catalog (SciCat Project 2017)) to store the data coming from these larger instruments and experiments *via* the RFI File Monitor.

RFI File monitor takes on a number of key tasks including looking at file systems for files being written and queueing up the data for data transfer to the object store and saving metadata to the metadata catalog (SciCat).

The goals of the Franklin Informatics Strategy are to make all Research Data findable, accessible, and reusable in line with FAIR principles, both within the institute and externally. Curlew Research undertook an institute-wide review of services in place that support good practice in managing research data, providing a snapshot of current practice at the Institute, and providing a benchmark against which development can be measured in the coming years.

## The Approach Framework
### The UKRI Concordat, the FAIR Cookbook and the Dataset Maturity Model

The UKRI Concordat on Open Research Data defines 10 Principles to ensure that data gathered and generated is properly managed, and made accessible, intelligible and usable by others (Figure 1.). Particular relevance for the development of data management governance and processes at the Franklin is given to the principles that state that open data are an enabler of high-quality research and a facilitator of innovation, that good data management be established at the outset, and that data curation is vital to enable reuse both by the originator and by others, and these principles formed the focus of this review.

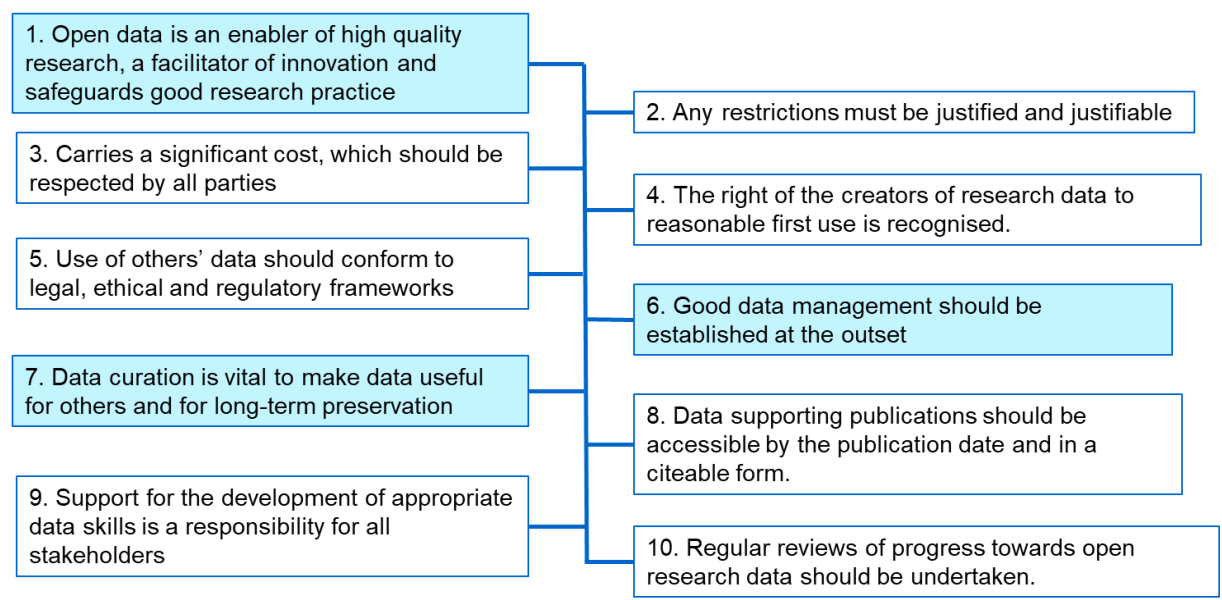

**Figure 1. The 10 Principles of the UKRI Concordat on Open Research Data.** Particular relevance for the Franklin is given to the principles highlighted in blue, that state that open data is an enabler of high-quality research and a facilitator of innovation, that good data management be established at the outset, and that data curation is vital to enable reuse by others.

Early in its lifecycle FAIR provided high level principles for making research data Findable, Accessible, Interoperable and Reusable and identified 15 guiding principles for making data FAIR. The original FAIR work provided this high-level model and since then groups have been working on developing approaches to the FAIRification journey. One of these projects is the IMI FAIRplus project (ELIXIR 2022) which has been developing a series tools and guidelines to support effective FAIR workflows. The FAIRplus project has provided detailed methodology for assessing the FAIRness of Research data and provides guidance and tools for the processes needed for FAIRification of data in the Life Sciences, in the form of the FAIR cookbook (Rocca-Serra *et al*. (2022) – a collection of use cases and over 70 recipes for making and keeping data FAIR. A generalised process for the FAIRification of research data is shown in Figure 2. Steps in the process include (i) describing use cases through discussion with domain experts and identifying business questions or the purpose of reusability; (ii) analysing the datasets: assessing the current level of FAIRness, identifying concepts and entities and assigning Identifiers; (iii) identifying appropriate ontologies to describe the data, (iv) mapping the data: creating URIs, annotating entities and editing a metadata catalog; (v) creating a distribution endpoint (such as an Institution Data Registry and Catalog) and defining how access to data is achieved; the whole FAIR strategy should be integrated with the analytical environment, resulting in ease of use and normalisation of the FAIRification process within the workflows. In the Franklin Data Management Review we aimed to assess the current levels of FAIRness of research data, focusing on how the core principles Findability, Accessibility and Reusability of data are currently met at Franklin.

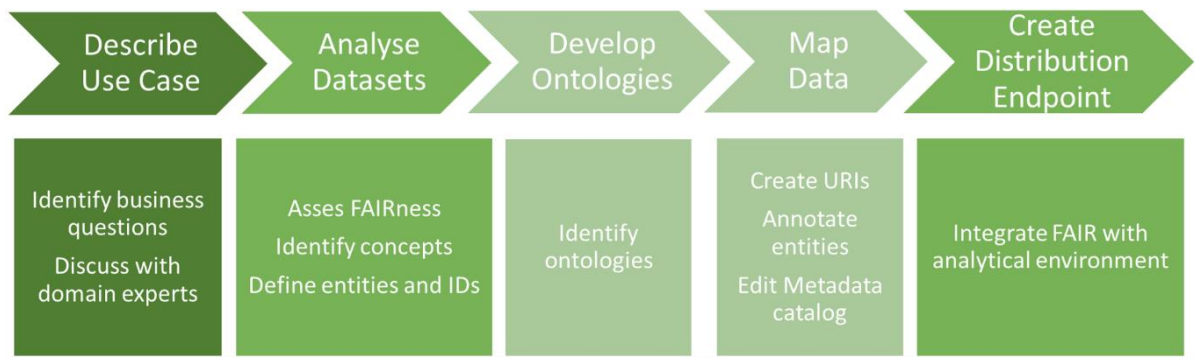

**Figure 2. Generalised Process Flow for the FAIRification of Research Data.** Steps in the process for FAIRification of research data include describing use cases, analysing the datasets, assessing the current level of FAIRness, identifying appropriate ontologies to describe the data, mapping the data: creating URIs and annotating entities, editing a metadata catalog, creating a distribution endpoint and defining how access to data is achieved, and integrating the whole FAIR strategy with the analytical environment (adapted from Grebe (2021)).

The implementation of each of the 15 FAIR principles and challenges that may arise with each are discussed in Jacobsen *et al.* (2020).

A key element of the journey for FAIRification is assessing the starting point and the steps required to achieve the desired outcome. The FAIRplus-DSM model (Beyan 2021) describes five levels of dataset maturity, characterised by increasing attainment of the FAIR requirements across three categories:

- **Content**: what is reported in the dataset & the metadata.
- **Representation & format**: how the data object & metadata object are represented and formatted.
- **Hosting environment capabilities**: what are the capabilities of the hosting environment to enable and support the use of FAIR data.

The Maturity Levels are defined further in the documentation, along with Indicators of attainment based on previous work around the FAIR indicators, undertaken by the Research Data Alliance (Herczog *et al.* (2020) and the FAIRsFAIR (Deveranu *et al.* (2020) projects.

In this initial work with the Franklin we focused on how the core principles Findability, Accessibility and Reusability of data are currently met at the Franklin. Since the fourth principle, Interoperability, can prove more difficult to implement and requires a definition of purpose, this will be considered in a later phase of the analysis and review process. We considered how data are currently captured and archived, and how the research data may be discovered and accessed by researchers both within project groups and by other researchers external to the Institute.

## Methodological Approaches for the Review

Curlew Research work with organisations that include pharma, biotech and research institutes, to review their data management strategies. We adopt a number of methods when conducting a data management review, including conducting stakeholder interviews, running workshops with scientists, and conducting online surveys in order to gather views from a wider number of researchers across the organisation (Figure 3.).

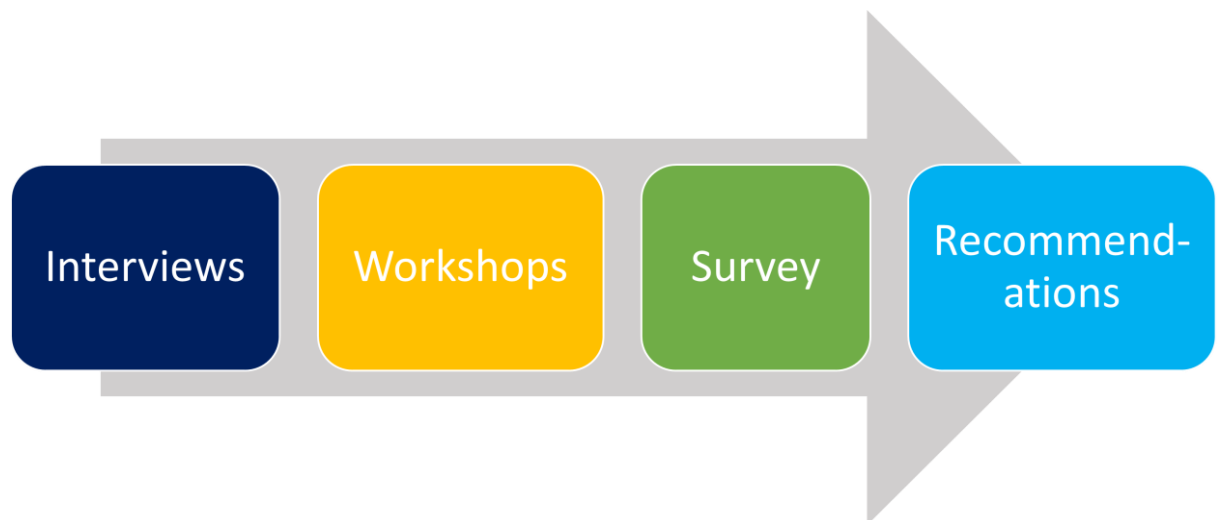

**Figure 3. The Methodological Approach for a Data Management Review adopted by Curlew Research**

### Interviews

We create a set of personas to support interviewee selection and mappings to life science workflows and data management. We identify key personas (domain experts) related to the activities and lifecycle of data within the organisation which include:

- data creators (e.g. project scientists)
- data owners (e.g. Project Leads)
- data stewards
- data scientists
- collaborators plus other roles where appropriate

Starting the review process with interviews allows gathering of important background information and the broader strategy of the organisation. Interviewing individuals as opposed to a group interview allows the interviewee space to express opinions. We consider the time spent on each technique proportionate to the value of information gained, and although time consuming, with interviews lasting one hour, we consider interviews an important part of the process. Example interview questions include role descriptions, the nature of distribution of researchers in the group – *e.g.,* at the Institute, at Oxford University, at Diamond, other Universities, the nature of collaborations with other Institutes and Industry, gap analysis and vision for the future.

### Workshops

Working with domain experts in a series of in-person workshops describing use cases, we identify the purpose(s) of reusability of data and analyse the datasets to assess the current level of FAIRness. We ask researchers to document their workflows in a manner similar to the management visualisation tool SIPOC, which asks users to consider the workflow in terms of **s**uppliers, **i**nputs, **p**rocesses, **o**utputs and **c**ustomers.

Having been shown a high-level example (Figure 4.), researchers are asked to chart their own workflow as a series of steps and map onto it the data existing at each stage. We ask researchers to identify concepts and entities within their workflows and discuss what metadata is needed to describe these. The data lifecycle elements can be further broken down to include instruments used to capture data and data processing tools at this early stage. Other critical elements related to data management are the data storage used at different points and the metadata used in cataloguing datasets and these were added to the workflow. In terms of differentiating these different elements, we use different coloured sticky notes to represent the elements. Further annotation can be added depending on the workflow being reviewed - for some, larger compute is needed and that can be shown; equally, collaboration elements can be shown where data needs to be shared with colleagues within or external to the Institute. All this helps build up the visual picture of the workflow with a focus on the data lifecycle and supporting elements

We consider both the capabilities and uptake (ease of use) of the processes already in place for data archiving and cataloguing that support FAIR principles.

Researchers are then asked to perform a gap analysis, identifying what is currently working well and what less well and to prioritise specific items for action. We adopt a simple voting system, one of the simplest being the use of red dots to allow attendees to vote with each person having 3 votes to use users are also asked the open question 'In an ideal world what would make your workflow easier?'

The use of brown paper, "sticky" notes and coloured dots for prioritisation in this exercise and the act of 'building something' encourages contribution and gives participants the opportunity to learn from each other. This workshop exercise can act to prime the researchers involved for later discussions which bring scientists from different disciplines together.

One benefit to this approach is that other ideas and elements surface in the workshop. Depending on time and focus these new ideas can be discussed in the workshop, Open discussion of these new ideas can take the workshop in directions not expected but is often really valuable in allowing issues and understanding to be shared. Alternatively, these new issues can be the subject of future workshop or issue analysis.

In the workshop we sought to balance the giving of information on the importance of good data management with the kinaesthetic learning style of the activity. Introducing information can be useful, but allowing the scientists to express their challenges and then relating this back to data management was more valuable. Our recommendation is to split the workshop hands-on activity into 2, with discussion and a regrouping of ideas at the half way point before introducing the latter parts of the exercise. For example if activity 1 is modelling the workflow and identifying data and how it is currently stored, we then introduce the concept of data sharing within and external to the institute, and ask participants to identify where data needs to be shared and with whom, and how would they currently search for and access the data. This avoids overloading the participants with information at the start of the workshop. Each workshop session was 90 mins of which 50 minutes was allocated to the workflow modelling exercise. Anything longer is big commitment to ask of researchers and 90 mins can be fitted around other experimental work for the participants.

It is also necessary to maintain an awareness of managing the unexpected. For example, there can be very different workflows within the audience e.g., for cryo-ptychography where the work is focused on development of technique and development of data processing methods compared to an experimental workflow for a protein crystallography experiment as described in Figure 4. It was possible to address both by maintaining a flexible approach and coordinating parallel sessions within the one workshop.

The workshops are especially effective in person. For other customers we have also run this exercise remotely using an electronic whiteboard with the option of adding sticky notes. However, we have found that, in online sessions, the element of free-flowing discussion is missing and participants tend to complete the tasks as individuals. Online sessions, therefore, necessitate a more structured activity with repeated opportunities to regroup and discuss points raised by individuals.

## Data Management Survey

In order to gather views from a wider number of researchers across the organisation, we also design and circulate a questionnaire where we ask scientists across all domains for their views on, for example, data processing and archiving, metadata, business rules and Informatics support. The survey is anonymous and takes around 15 minutes to complete. We ask respondents to describe the nature of their research and the size of datafiles and frequency of data collection. Responses for categorical questions are presented as either yes/no radio buttons or a checkbox list asking the user to tick all that apply. For every question we also provide a free text box for optional additional comment, Questions can also be designed to capture an emotional response on a Likert scale (*e.g.,* I strongly disagree to I strongly agree). Values for all Likert questions are normalised so that 1 always represents a poor outcome (*e.g.,* the inability to archive raw data, or lack of understanding) and 5 always represents a good outcome (*e.g.,* the ability to archive raw data or good understanding of a process). When plotted on a percentage scale as diverging stacked bar charts, they highlight areas for future focus across the organisation (Figure 5.).

Responses to Likert questions may also be plotted as radar plots allowing comparison of satisfaction for the individual domains, highlighting where domain-specific data management solutions might benefit researchers (Figure 6.). The survey questions and format are provided in the Supplementary Information.

The online questionnaire relies heavily on getting engagement from the audience and achieving a reasonable number of replies from the audience. We have found that using an internal sponsor for the work is helpful in encouraging participation, as is showing changes that have come about from the previous questionnaire iterations. In some organisations there can be a case of survey fatigue and so one needs to be mindful of how often these efforts are run and respectful of peoples' time. Many organisations offer the chance to win a prize in a draw as encouragement for completion of a survey. This needs to be managed carefully while maintaining the anonymity of respondents.

## Outputs

All the information obtained from the interviews, workshops and survey is analysed by Curlew Research and distilled into a final report which provides a snapshot of current practice for the organisation. The report summarises the data management infrastructure and intended practice currently in place at the organisation and highlights where users find this easy to incorporate as part of their normal workflow. The analysis includes an ordered list, by scientific domain, of areas that were already successfully implemented, and the researchers' priorities for action. These insights, together with the assessment of the current FAIRness of data, enable Curlew Research to make key recommendations for developing the Informatics and Data Storage Strategy.

# The Research Data Management Review for the Rosalind Franklin Institute: a case study

The Rosalind Franklin Institute wished to review its data management approaches and associated platforms in the context of the UKRI Concordat on Open Research Data. The institute-wide analysis, carried out by Curlew Research using the methodologies described above, provided the Franklin with material to support future data governance planning and activities, and support both the needs of the key stakeholders within the Franklin, but also key collaborators, both internal and external. The Review was set up through an appropriate legal and commercial agreement between the Institute and Curlew Research to help manage privacy and consent issues. Anonymity of all participants was preserved. Ownership of the material, findings and report remain with the Franklin.

## Interviews

Discussions on Informatics Strategy were held with the Head of the AI and Informatics, and with the Senior Research Software Engineer who has a role in developing the Data Archiving and Catalog Infrastructure (RFI File Monitor and SciCat) at the Franklin.

We conducted online interviews with Theme Leads for four out of the five Research Themes at the Franklin (Structural Biology, Correlated Imaging, Next Generation Chemistry and Artificial Intelligence and Informatics (the fifth, Biological Mass Spectrometry, being a new emerging Theme at the Franklin)). We followed these with online interviews with three researchers within the Structural Biology Theme.

This initial analysis of the Franklin workflows, covering key internal capabilities and external collaborations, set the scope for the Data Management Workshops subsequently held with researchers in each of the three Scientific Themes and we identified a number of areas for focus for the analysis through conducting the interviews:

- Data Capture and Archiving for the larger instruments
- Data Findability as key for FAIR principles alignment
- Data Processing (larger instruments) and Analysis
- Data Management within the 'wet lab'
- Metadata
- Data Sharing

## Workshops

Researchers from three of the Scientific Themes took part in the Workshop exploring data management within the following workflows:

- Correlated Imaging
    - Experimental design and data collection for Electron Microscopy studies
- Structural Biology
    - a Protein Crystallography study

- a Tomography workflow
- Next Generation Chemistry
  - a protein functional study, integrating the Mass Spectrometry and NMR analysis workflows

The Workshop started with a discussion on the value of Open Research Data – why we should be aiming for Open Data and who stands to benefit, an introduction to the FAIR concepts of Findability, Accessibility, Interoperability and Reusability, and the concept of metadata. Following this, and having been shown a high-level example (Figure 4.), researchers were asked to chart their own workflow as a series of steps and map onto it:

- the data generated at the start of a study *e.g.,* publications, data on protein structure and function
- the initial data for samples coming into the lab and how this is received and stored
- entities and assets existing at each stage and how they are identified
- the data generated at each step
- how the data is currently captured and archived
- how the data is currently retrieved and processed
- what metadata is generated throughout the study and how this is captured
- how the data is shared, both within the project group and with external collaborators

We considered both the capabilities and uptake (ease of use) of the processes in place for data archiving and cataloguing (*e.g.,* RFI File Monitor, SciCat, file storage and computational processing power) that support the Franklin's Open Research Data policy.

Researchers were then asked to identify what was currently working well, and what less well and to prioritise specific items for action. They were also asked the open question 'In an ideal world what would make your workflow easier?'

The workshops proved valuable in priming the audience for further discussions held with the Informatics management team and involving researchers from all scientific Themes.

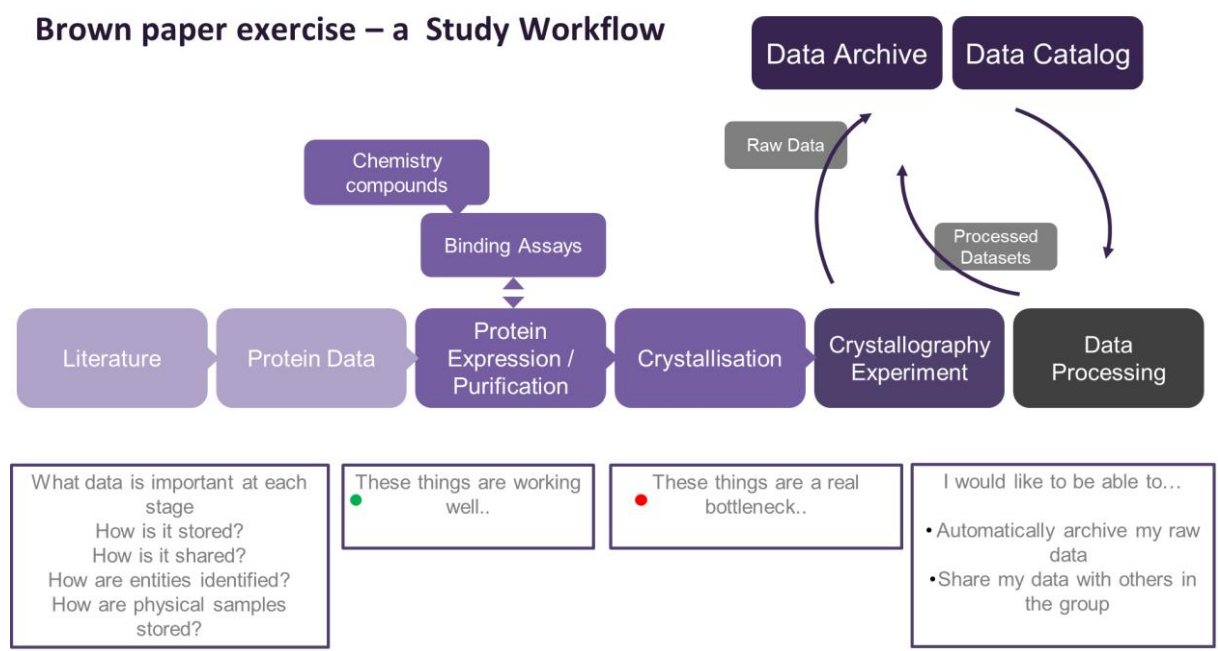

**Figure 4. High Level workflow for the Workshop.** Researchers were asked to chart their own workflow as a series of steps and map onto it the data flow in the study.

## Data Management Survey

In order to gather views from a wider number of researchers at the Franklin, we also designed and circulated a questionnaire where we asked scientists across all Research Themes for their views in the following five areas:

- Data Capture and Archiving for the larger instruments
- Data Management within the 'wet lab'
- Data Processing
- Metadata
- Informatics Support at the Franklin

For the survey we made use of different question formats which included the (multi) selection of responses describing, for example, how data was archived, and free text comment. Three of the areas were the focus of 16 questions for the Likert analysis (Data Capture and Archiving, Data Processing and Informatics Support).

The survey, created using Microsoft Forms, was circulated to all Franklin-based researchers with a request for completion. 22 researchers responded to the survey (a 22% response rate) with all five Research Themes represented.

The views across all Research Themes for the 16 Likert questions were plotted on a percentage scale as diverging stacked bar charts, highlighting areas for future focus across the Institute (Figure 5.).

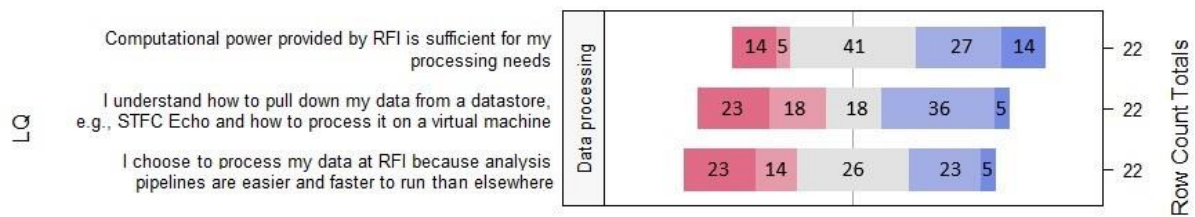

**Figure 5. Survey Responses.** Responses to Likert questions plotted as diverging stacked bar charts allow the best visualisation of this type of data. Numbers indicate the proportion of responses for each satisfaction rating as a percentage of the total responses for that question.

In order to assess the satisfaction rating by Research Theme for each of the three areas which were the focus of the Likert analysis (Data Capture and Archiving, Data Processing and Informatics Support) responses to Likert questions were also plotted as radar plots (Figure 6.). This allowed comparison of satisfaction for the individual Research Themes. It was apparent that newer research groups, recently arrived at the Franklin, were less satisfied overall with Informatics support and the services in place for data management, than were the more established research groups at the Franklin, highlighting where further effort in communication and training, or theme-specific data management solutions might benefit researchers.

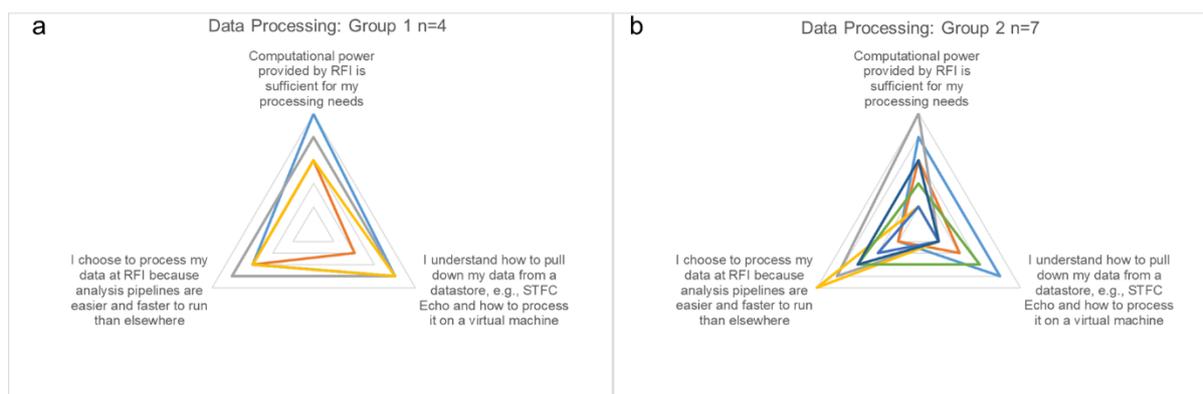

**Figure 6. Survey Responses.** Responses to Likert questions for individual Research Themes plotted as radar plots. Colour denotes the responses by individual participants, with the satisfaction for individual questions rated from 1 (strongly disagree, innermost polygon) to 5 (strongly agree, outermost polygon). Thus, a more open polygon denotes a higher overall satisfaction score within a Research Theme, and the radar plots provide visualisation for comparison of satisfaction between different Research Themes. For the questions asking about Data Processing, it is evident that a more established group, Group 1 (a), was more satisfied overall than a more recently arrived Group 2 (b).

Researchers were also asked to comment on data file size and frequency of generation in the survey. As might be expected with such wide-ranging research and analytical techniques, data file size and frequency across the Franklin varies enormously. For example, for datasets for single particle Cryo-EM of membrane proteins, the workflow covers everything from constructs in different expression vectors (mostly for expression in bacteria), purification of proteins (gels, chromatograms) and data collection. Raw data are movies from a cryo- electron microscope, data processing is carried out to obtain a map, and model fitting is performed to obtain the structure. Typically, tens of thousands of files are generated in an experiment and for the entire workflow multiple TB data are collected with file sizes ranging from KB to GB. The frequency of collection varied, for 'wet lab' experiments data is collected every week, for the Cryo-EM experiments data collection is every month/every couple of months.

- Within AI and Informatics, datasets for Neural Network analyses on tomography data can generate ~ 66GB data, with several individual files over 1GB.
- Within Biological Mass Spectrometry, high dimensional imaging datasets of over 100GB are created 3-4 times a week, consisting of one large data file plus multiple small metadata files.

These ranges of file sizes and file number (volume) create challenges for any data management infrastructure. This should encourage thinking about the different combinations *e.g.,* large file frequency & average file size versus average file frequency & large file size to assess whether different pathways need to be put in place, plus options for asynchronous file transfer and batch working.

### Outputs

The main output from the analysis was the final report produced for the Franklin which provides a snapshot of current practice for the Institute. The report was written as an internal review and currently remains internal to the Franklin. The aim is to share learnings over time.

The report summarises the Informatics Infrastructure and intended practice currently in place at the Franklin and highlights where users found this easy to incorporate as part of their normal workflow. The assessment of the current FAIRness of data resulted in key recommendations for developing the Informatics Strategy over the next years.

Appendices to the main report included

- An ordered list, by Research Theme, of
    - areas that were already successfully implemented
    - the Researchers' priorities for action, separated into Informatics Priorities (such as automated data archiving) and IT Priorities (such as installation of ethernet or desktop software in instrumental computers).

- A Summary of Data File Size and Frequency of Data Collection in research across the Franklin taken from the survey.

## Audience Feedback

We received feedback that the workshops were valuable in both providing actionable insights for data management and in priming the audience for further discussions held with the Informatics management team and involving researchers from all scientific Themes. All researchers given the opportunity to feed back if they had further comment or strong disagreement - Curlew Research sent out emails to all participants thanking them for their contribution and requesting further comments. For this first review, we received no strong views and nobody brought up anything specific. Requests for feedback are integral to the review process and we would expect more feedback, reflecting on changes since the last review, when carrying out the review in subsequent years.

## Moving Forward

There is an impressive infrastructure for instrumental data capture and archiving in place at the Franklin. They are now focused on increasing awareness through continued training, ease of use, and normalisation of FAIRification practices in the analytical workflows.

A key piece of information to come out of the Workshops was that while the platforms and intended practices for making data Open (*e.g.,* automated data capture, cataloguing, and provision for open access) are well-developed for data coming from the 'large instruments', project data arising in the 'wet lab' was less well described or captured. The report recommended effort should focus on FAIRification of project data, as described in Figure 1. In particular, the identification of entities (*e.g.,* proteins, expression systems, grids or crystals, chemical compounds), the creation of persistent identifiers, and the use of structured metadata and ontologies to describe these will enable the Findability of data in a research context. The use of a common format for experiment capture including electronic laboratory notebooks (ELN) will enable data capture and sharing within project groups and will be included in the next phase of the Informatics strategy.

## Concluding Remarks

FAIRification and making research data open should be viewed as a journey, with a long-term view of improving the potential for reuse of research data. It requires investment in technical solutions but also in stewardship and curation resources.

It also requires a cultural shift in favour of good research data management practice among researchers who traditionally have a view of owning their own data. In part this is met by the requirement of leading Journals that data be made open prior to publication of research. Usually this is achieved by data deposition in public repositories created specifically for various scientific disciplines, *e.g.,* EBI/EMBL EMPIAR, EMBD, PDB. However, normalisation of practice as well as ease of use and embedding of data FAIRification tools in analytical workflows are necessary in order for research data to be made FAIR at the point of creation – 'born FAIR', and thus easily made accessible as appropriate.

User-driven engagement through the workshops held in this review resulted in key actionable insights including:

- Continuing increased engagement with research scientists resulting in understanding the challenges faced, and leading to a shared vision for Open Data for the Franklin
- Identifying where scientists would benefit from targeted training
- Increasing contact with some of the scientific groups allowed identification of where, with limited resources, changes can have most impact
- The evolution of mechanisms for data transfer, focusing first on data moving, and then on data sharing. This involved restructuring data access in terms of identifying
    - a data owner – the scientist who ran the experiment
    - a data curator – the scientist running the instrument
    - data access groups – defined around themes and collaborations
- Identifying the necessity for Business Rules for data sharing and backup

The survey provided both a baseline to build on – a general 'state of the Nation' relating to the Concordat – by which progression can be measured year-on-year, and the opportunity to explore in detail the take up of intended data management practices across the Institute. Rerunning the survey on a yearly basis will inform data management plans. Principle 10 of the UKRI Concordat on Open Research Data states that regular reviews of progress towards open research data should be undertaken, and that 'this should be manifested in the undertaking of regular reviews that monitor progress and register issues to be addressed' and that 'Such reviews should not be over-burdensome but rather flexible and recognise that developments will take time. Their essence should be one of identifying and sharing best practice.'

This initial assessment provided benchmarking by which progress towards Open Research Data can be measured in the coming years, and recommendations provided for the short and medium term. Data management should be aligned to the maturity of the organisation and growing needs for data re(use). Data management plans should be reviewed and updated on a regular basis in light of this maturity. One way to show the progress of Data Management is through the definition and then the monitoring of certain key indicators - here the adoption of the recommended processes in place for data archiving and sharing. For this type of analysis, we recommend a yearly gap between the assessments to allow time for changes to be made and rolled out to the user community and for that impact to be recognised. As the organisation matures and there is less change then this time period could extended, but such is the pace of change with new instruments and techniques that a yearly frequency allowing focus is appropriate at the moment.

## Competing interests

FC and NL completed paid consultancy work for Curlew Research as part of the data acquisition for this study. All other authors have no competing interests.

## References

Baskerville (2021) Baskerville A national accelerated compute system, https://www.baskerville.ac.uk/ [Last accessed 13 March 2023]

Beyan, O., Emam, I., Rocca-Serra, P., Sansone, S-A., Juty, N., Alharbi, E., Wood, C., Henderson, D., Burdett, T., & Konopko, M. (2021). D2.5 FAIRplus FAIR Data Maturity Framework. Zenodo. https://doi.org/10.5281/zenodo.5040592

Devaraju, Anusuriya, Huber, Robert, Mokrane, Mustapha, Herterich, Patricia, Cepinskas, Linas, de Vries, Jerry, L'Hours, Herve, Davidson, Joy, & Angus White. (2020). FAIRsFAIR Data Object Assessment Metrics (0.4). Zenodo. https://doi.org/10.5281/zenodo.4081213

Diamond Light Source, https://www.diamond.ac.uk/ [Last accessed 13 March 2023]

ELIXIR (2022) IMI FAIRplus Project, https://fairplus-project.eu/ [Last accessed 13 March 2023]

Grebe, A., Keynote Presentation, 2nd FAIRplus Innovation and SME Forum (2021) https://drive.google.com/drive/folders/13b4CUSADxJsMKtw2Qg1P50pwMNcyPZz4 [Last accessed 13 March 2023]

Greeves, E. (2020) Echo: A Ceph-backed storage service for particle physics data, https://www.scd.stfc.ac.uk/Pages/Echo-leaflet.aspx [Last accessed 13 March 2023]

Herczog, E., Russel K., Stall S. (CoChairs) (2020) FAIR Data Maturity Model. Specification and Guidelines (1.0). Research Data Alliance FAIR Data Maturity Model Working Group. https://doi.org/10.15497/rda00050

Jacobsen, A., de Miranda Azevedo, R., Juty, N., Batista, D., Coles, S., Cornet, R., Courtot, M., Crosas, M., *et al.* 2020. FAIR principles: interpretations and implementation considerations. Data Intelligence, 2(1-2), pp.10-29. https://doi.org/10.1162/dint_r_00024Rocca-Serra, P., Sansone, S-A., Gu, W., Welter, D., Abbassi Daloii T., & Portell-Silva, L., (2022) D2.1 FAIR Cookbook, Zenodo, https://doi.org/10.5281/zenodo.6783564

Schoonjans, T. (2021) RFI File Monitor, https://rosalindfranklininstitute.github.io/rfi-file-monitor/ [Last accessed 13 March 2023]SciCat Project (2017) https://scicatproject.github.io/documentation/ and https://github.com/rosalindfranklininstitute/ [Both Last accessed 13 March 2023]

Science and Technology Facilities Council Open Stack Cloud, https://www.scd.stfc.ac.uk/Pages/STFC-Cloud-Operations.aspx [Last accessed 13 March 2023]

Science and Technology Facilities Council, Scientific Computing Services, https://www.scd.stfc.ac.uk/Pages/Services.aspx [Last accessed 13 March 2023]

UKRI (2016) UKRI Concordat on Open Research Data -, [https://www.ukri.org/wp-content/uploads/2020/10/UKRI-020920-ConcordatonOpenResearchData.pdf](https://www.ukri.org/wp-content/uploads/2020/10/UKRI-020920-ConcordatonOpenResearchData.pdf) [Last accessed 13 March 2023]

UKRI (2023), [https://www.ukri.org/manage-your-award/publishing-your-research-findings/making-your-research-data-open/](https://www.ukri.org/manage-your-award/publishing-your-research-findings/making-your-research-data-open/) [Last accessed 13 March 2023]

Wilkinson, M., Dumontier, M., Aalbersberg, I., Appleton, G., Axton, M., Baak, A., Blomberg, N., Boiten, J-W. *et al.* (2016) The FAIR Guiding Principles for scientific data management and stewardship. Sci Data 3, 160018. https://doi.org/10.1038/sdata.2016.18